\documentclass[%
 reprint,
 amsmath,amssymb,
 aps,
]{revtex4-2}
\usepackage{subfigure}
\usepackage{graphicx}% Include figure files
\usepackage{dcolumn}% Align table columns on decimal point
\usepackage{bm}% bold math
\usepackage{siunitx}
\usepackage{appendix}
\usepackage{amsmath,array}
\usepackage{kantlipsum}
\usepackage{float}
\usepackage{multirow}
\usepackage{adjustbox}
\usepackage{booktabs}
\usepackage{stackengine}
\begin{document}

\preprint{APS/123-QED}

\title{Quantifying and mitigating optical surface loss in suspended GaAs photonic integrated circuits}% Force line breaks with \\

\author{Robert Thomas$^1$}
\author{Haoyang Li$^1$}
\author{Jude Laverock$^2$}
\author{Krishna C. Balram$^{1, }$ }%
 \email{krishna.coimbatorebalram@bristol.ac.uk}
\affiliation{[1] Quantum Engineering Technology Labs and Department of Electrical and Electronic Engineering, University of Bristol, BS8 1UB, UK}
\affiliation{[2] School of Chemistry, University of Bristol, Cantocks Close, Bristol BS8 1TS, UK}

\begin{abstract}
Understanding and mitigating optical loss is critical to the development of high-performance photonic integrated circuits (PICs). Especially in high refractive index contrast compound semiconductor (III-V) PICs, surface absorption and scattering can be a significant loss mechanism, and needs to be suppressed. Here, we quantify the optical propagation loss due to surface state absorption in a suspended GaAs photonic integrated circuits (PIC) platform, probe its origins using X-ray photoemission spectroscopy (XPS) and spectroscopic ellipsometry (SE), and show that it can be mitigated by surface passivation using alumina (Al$_2$O$_3$). We also explore potential routes towards achieving passive device performance comparable to state-of-the-art silicon PICs.
\end{abstract}

\maketitle

%\tableofcontents

\section{\label{sec:level1}Introduction}

Compound semiconductors like GaAs, have superior electronic properties (for ex: electron mobility) compared to silicon \cite{del2011nanometre}. On the other hand, silicon has long dominated the microelectronics industry on account of the low surface and interface state ($D_{it}$) density at the Si-SiO$_2$ interface. Similarly in integrated photonics, despite silicon's limitations as an optical material (indirect bandgap, lack of a $\chi^{(2)}$ nonlinearity), it remains the dominant material platform. While the analogy between microelectronics and photonics is instructive, it is important to note that in photonics, there will always be a need for III-V materials for light emission. There are a variety of approaches currently being pursued from monolithic \cite{norman2019review} to hybrid integration \cite{liang2021recent} to get III-V materials on silicon photonic integrated circuits (PICs). But, given that all the important optical functionalities (light generation, fast modulation, routing, detection and signal processing) can be implemented in III-V semiconductors, it begs the question as to why the dominant trend has been hybrid integration of III-Vs on  silicon \cite{zhang2019iii}, rather than monolithic integration in a III-V platform \cite{smit2019past}?

While there are several factors that can explain this trend, one key determinant is whether monolithic III-V implementations can match the scale (footprint and component density) and performance (optical loss) of silicon PICs. Optical loss is especially critical as it underpins everything from laser linewidth \cite{santis2014high} to insertion loss of modulators \cite{zhang2021integrated}. Standard Si PIC offerings (silicon thickness $\sim$ 220 nm and 340 nm) have propagation loss $<$ 1 dB/cm. For monolithic integration in III-Vs to succeed, and the intrinsic material advantages of III-Vs to dominate, the propagation loss in III-V PIC platforms with comparable refractive index contrast (${\Delta}n>$ 1.5), and therefore component density, needs to be in the same range. In this work, our main focus is on mitigating the (linear) surface loss component which can be a significant source of loss in sub-\SI{}{\micro\meter} waveguide geometries. We focus on gallium arsenide (GaAs), the III-V material with the most established electronics foundry infrastructure \cite{del2011nanometre}, and therefore ideally placed to make the electronics-photonics transition like silicon. We would like to note that there have been exciting developments in low-loss photonics in related III-V material families like aluminum gallium arsenide (AlGaAs) \cite{xie2020ultrahigh}, gallium phosphide \cite{wilson2020integrated}, and indium phosphide \cite{jiao2020inp}. 

\begin{figure}[ht]
\centering
\includegraphics[width=\linewidth]{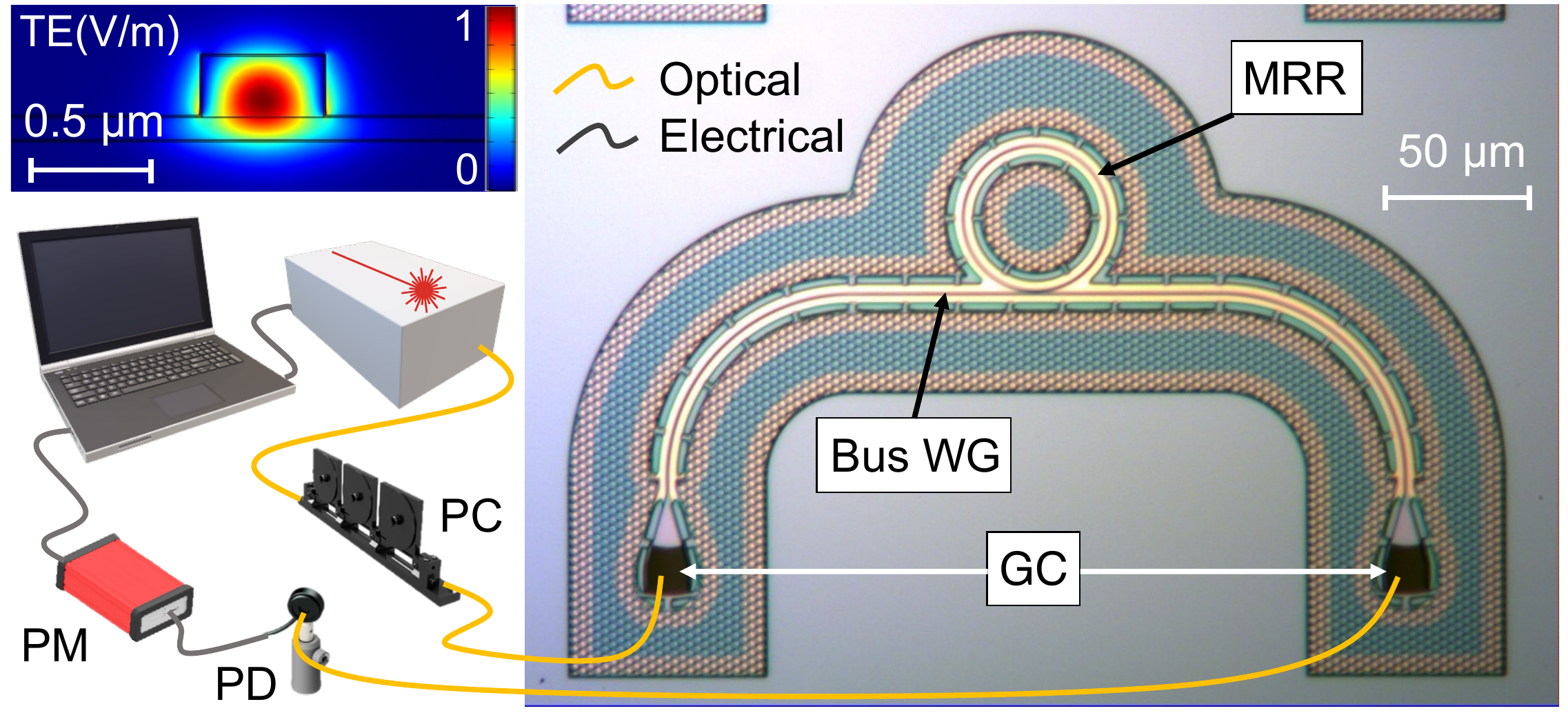}
\caption{Optical microscope image of a suspended GaAs PIC microring resonator (MRR) device with a schema of experimental setup (left). GC: grating couplers, PC: Polarisation controller, PD: Photodiode, PM: power meter. (top left) FEM simulation of the transverse electric field (TE) distribution of the fundamental mode of a suspended GaAs rib waveguide.}
\label{fig1}
\end{figure}

While there has been some recent work to address the issue of surface passivation in GaAs photonic devices \cite{guha2017surface, najer2019gated, stanton2020efficient, kuruma2020surface, jacob2022surface} building on work in GaAs electronics on developing MOSFETs \cite{ye2003gaas, gougousi2016atomic, kumah2020epitaxial}, the overall question of how far surface passivation can help improve passive device performance in a complex PIC process remains open. In this work, we use microring resonators (MRR) fabricated using our suspended PIC platform \cite{jiang2020suspended, khurana2022piezo}, as a means to quantify surface loss, and the effects of surface passivation. We compare the intrinsic quality factor ($Q_i$) of resonators measured with (treated) and without (untreated) an 8 \SI{}{\nano\meter} thick conformal surface coating of Al$_2$O$_3$ (alumina) deposited via atomic layer deposition (ALD), and show that the $Q_{i}$ can be improved by a factor of $1.4$. To probe the origins of this surface loss, we perform X-ray photoemission spectroscopy (XPS) on GaAs samples, pre and post passivation, and show that the results can be replicated using spectroscopic ellipsometry \cite{rauch2022model}, which provides a robust method to monitor surfaces inside the cleanroom. 

\section{Experimental results}

The MRR devices used in this work are fabricated on two identical 7 \SI{}{\milli\meter} chips taken from a $370$ nm thick GaAs device layer with 1 \SI{}{\micro\meter} thick Al$_{0.62}$Ga$_{0.38}$As buffer using a process similar to our previous work \cite{jiang2020suspended, khurana2022piezo}. Post waveguide release and sample cleaning, one of the chips (treated) is coated with an (8.0 $\pm$ 0.2) \SI{}{\nano\meter} thick layer of alumina deposited by ALD at a chamber pressure of 26.67 \SI{}{\pascal} and temperature of (300 $\pm$ 2)\SI{}{\degreeCelsius}. Figure 1 is a magnified image of a suspended GaAs MRRs used in this work, indicating the main photonic components of the device; grating couplers, bus waveguide and the micro-ring resonator. The loaded quality factors of the treated and untreated devices are calculated by fitting a Lorentzian \cite{gao2022probing} function to the micro-ring resonances, accounting for the asymmetry in the off-resonance transmission due to background Fabry-Perot resonances.  

\begin{figure}[ht]
\centering
         \begin{subfigure}
         \centering
         \includegraphics[width=0.98\linewidth]{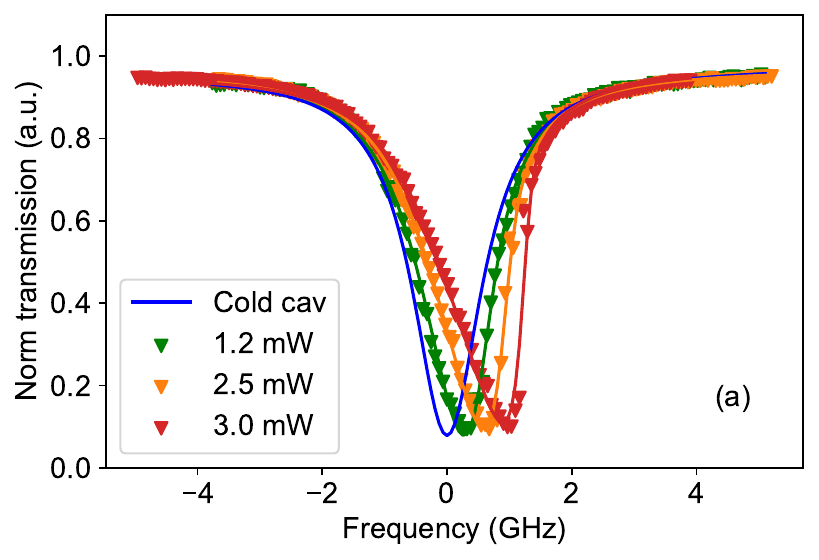}
         \end{subfigure}
         \begin{subfigure}
         \centering
         \includegraphics[width=0.98\linewidth]{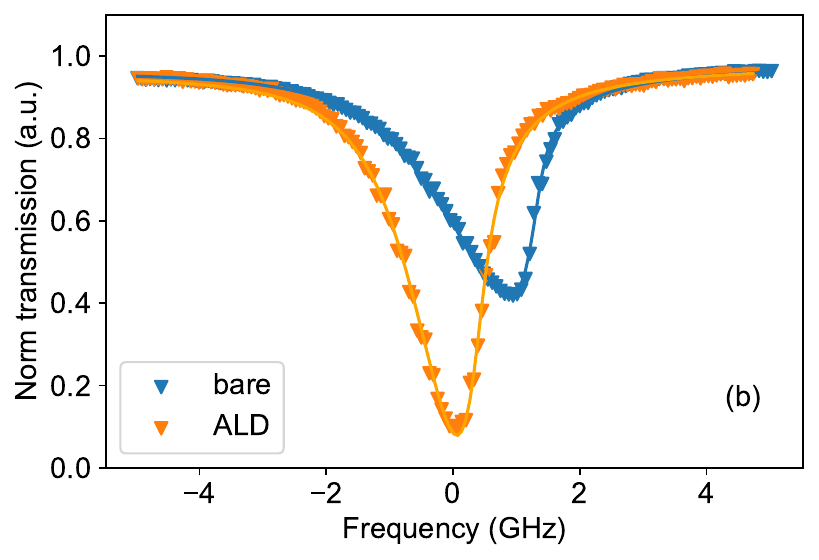}
         \end{subfigure}
\caption{(a) MRR transmission spectra measured for an ALD treated device at three different intracavity optical powers: 1.2 mW (green triangles), 2.5 mW (orange triangles) and 3.0 mW (red triangles) with fitted curves (matching coloured lines) and derived ‘cold cavity’ lineshape (blue line). (b) MRR transmission spectra measured for a bare untreated device (blue triangles and line) and an ALD treated device (orange triangles and line) at an intracavity optical power of (1.5 $\pm$ 0.1 mW).}
\label{fig2}
\end{figure}

Transmission spectra recorded from three different MRR devices with wavelengths that fall within $\pm$ 5 \SI{}{\nano\meter} of the centre wavelength of the grating coupler response are considered for both the treated and untreated samples, for a total of ten resonances for each case. High Q micro-resonators are susceptible to mode splitting of the resonant mode due to reflected light, either distributed reflections from surface roughness or point reflections from the coupling region \cite{li2016backscattering}. Resonant modes that exhibit a visible doublet splitting are not considered in the analysis due to the complexity of mode interactions at high intracavity powers, and the related challenge of robustly fitting the transmission spectrum. Figure 2a shows MRR transmission spectra for a treated device at three different intracavity optical powers. The characteristic shark-fin shape of the resonator transmission is due to absorption induced heating of the resonator which shifts the cavity frequency due to the thermo-optic effect \cite{barclay2005nonlinear, gao2022probing}. 

The thermo-optic effect can be modelled within a Lorentzian fitting function for the cavity transmission $T(\omega)$ \cite{gao2022probing}, by adding a term ($\beta$), which produces a resonance frequency shift proportional to the intracavity optical energy density, $\rho$. The measured resonance spectra are fitted numerically, starting at an off-resonance frequency, where $\rho = 0$, and iteratively feeding this solution forward and solving for the other fitting parameters; $\kappa_c$, the bus waveguide cross coupling rate, and $\kappa_T$, the total loss rate (the sum of the coupling loss rate and the intrinsic waveguide loss rate, $\kappa_i$, i.e. $\kappa_T = \kappa_c + \kappa_i$), at each step. 
   
\begin{equation} \label{eq:1}
T(\omega) = \left[ 1  - 
	\frac{\kappa_c}{ \frac{\kappa_T}{2} +              i(\omega_0 - \omega + \beta\rho)} \right]^2
  \end{equation}
  
By definition, the $Q_i$ of the resonator is proportional to the intrinsic waveguide loss rate: $Q_i = \omega_0/\kappa_i$, where, $\omega_0$, is the centre frequency of the cavity resonance. The intrinsic propagation loss is given by $\alpha_i = n_g/c\cdot{\kappa_i}$, where $n_g$ is the group index of the ring determined from the free spectral range of the resonant modes, $\Delta\lambda_{FSR}$, via $n_g = \lambda_0^2/({\pi}d){\Delta\lambda_{FSR}}$, where $d$ is the diameter of the MRR. $c$ is the speed of light in vacuum. 
 
The intracavity optical power i.e. the optical power circulating in the ring at the resonant frequency, is calculated by setting $\beta = 0$ in equation 1, thereby nullifying the thermo-optic frequency shift to produce a cold-cavity lineshape, as shown by the blue curve in Figure 2a. The finesse, $F$, and the coupling factor, $\gamma=1-\sqrt{T_0}$,  are extracted from the measured transmission data, with $T_0$ corresponding to the normalized cavity transmission at resonance. The optical power in the bus waveguide, $P_0$, is taken from the measured off resonance transmission value, accounting for the insertion loss (typical value: 5.0 $\pm$ 0.1 dB) of the input and output grating couplers. The MRR intracavity optical power can then be estimated as: $P_{cav} = {F/\pi}\cdot{\gamma}\cdot{P_0}$. Figure 2b shows the transmission spectra of an ALD device compared to a bare device at the same intracavity optical power 1.5 $\pm$ 0.1 mW. The extinction ratio difference between the resonances results mainly from the alumina cladding increasing the waveguide resonator coupling compared to the bare devices. This is accounted for implicitly in the model via the $\kappa_c$ term. From independent measurements of resonator quality factor as a function of resonator length with a fixed coupling geometry (Figure 7, Appendix A), we can determine that both the treated and untreated samples operate in the undercoupled regime.

The estimated $Q_i$ values for the ALD treated (orange) and bare untreated (blue) device resonances (ten each), measured with an intracavity optical power of 1.3 $\pm$ 0.4 mW  are shown as a histogram in Figure 3a. Statistical analysis indicates that the samples are normally distributed with the mean $Q_i$  values of the ALD treated and bare untreated sample groups, (2.2 $\pm$ 0.2) x 10$^5$ and (1.6 $\pm$ 0.2) x 10$^5$ respectively. The mean propagation loss, $\alpha_i$, values for the treated and untreated samples are thereby estimated to be 3.2 $\pm$ 0.4 dB/cm and 4.5 $\pm$ 0.6 dB/cm respectively. The alumina surface passivation has therefore reduced the surface loss in these devices by $\approx$ 1.4$x$.

\begin{figure}[ht]
\centering
\begin{subfigure}
         \centering
         \includegraphics[width=0.98\linewidth]{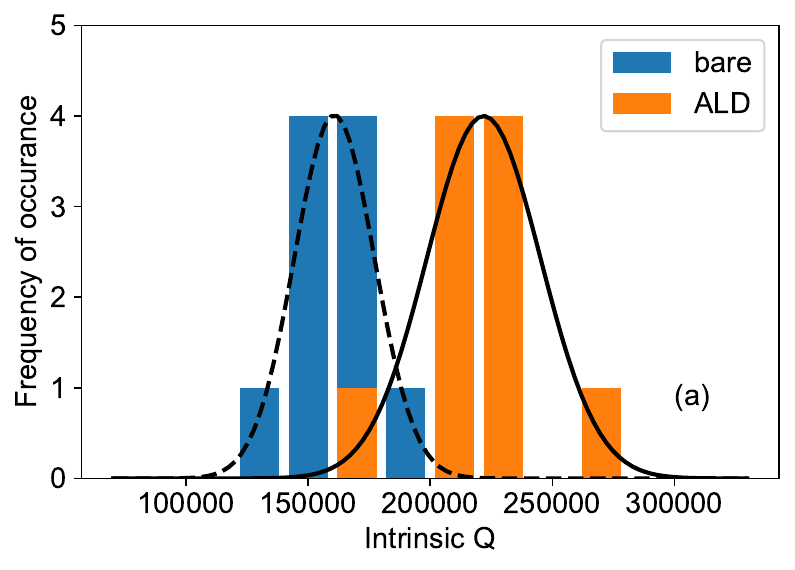}
         \end{subfigure}
         \begin{subfigure}
         \centering
         \includegraphics[width=0.98\linewidth]{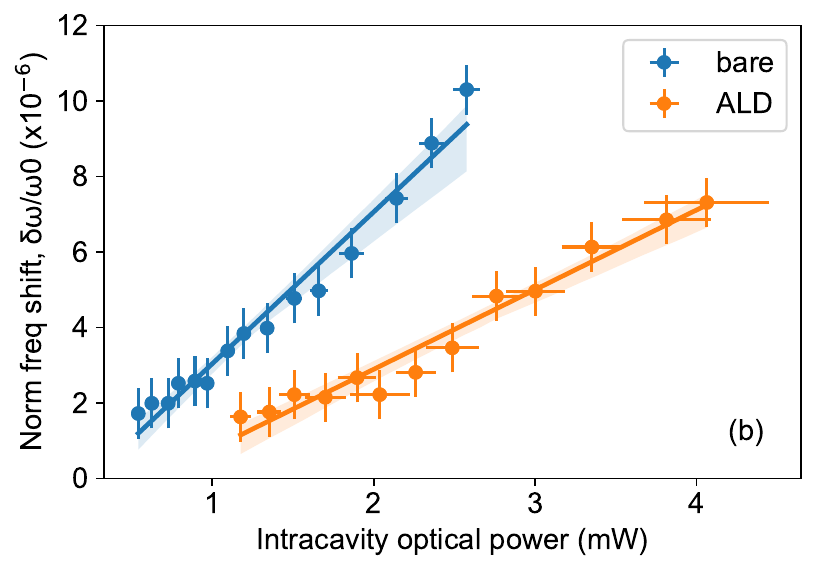}
         \end{subfigure}
\caption{(a) Histogram of measured $Q_i$ values from ten resonances each from bare (blue bars) and ALD treated (orange bars) MRR devices. (b) Normalized cavity resonance frequency shift as a function of intracavity optical power for the median $Q_i$ resonances of the untreated (blue circles) and treated (orange circles) sample groups.}
\label{fig3}
\end{figure}

Alumina surface passivation can reduce surface loss by reducing both the roughness induced scattering (by reducing the effective index contrast ${\Delta}n$) and by also reducing surface state absorption. Given that these samples are patterned using ebeam lithography, the waveguide surface roughness is not the main contribution to surface loss. 
We can confirm this by comparing the intracavity heating effects in the bare and treated devices. Figure 3b shows the normalised cavity resonance frequency shift for both ALD and bare devices, with median $Q_i$, plotted as a function of intracavity optical power. The lower limit of the power range is determined by our  tunable laser source (-14 dBm), and the upper limit is set by the onset of non-linear absorption effects. It can be clearly seen that the slope for the bare sample is significantly higher showing that the ALD does indeed reduce surface absorption, and therefore cavity heating. We assume that the thermal time constant of the device is not modified significantly with the ALD layer, although this needs to be verified in practice \cite{wang2021using}. Further details of the optical measurement analysis are included in Appendix A. 

To confirm the surface chemistry behind the absorption reduction, we carried out XPS measurements on the GaAs devices. Test samples for XPS are taken from a wafer with a similar layer structure (340/1000 GaAs/AlGaAs) to that of the MRR devices, but grown on a conductive N$^{++}$  doped substrate to avoid sample charging effects. XPS probes the upper surface of the sample to a depth of $\approx$ 10 \SI{}{\nano\meter}. The measurements were performed using a monochromatic Al K$\alpha$ X-ray source and the samples were degassed under ultra-high vacuum at 350 $^{\circ}$C for 1 hour. XPS of the Ga and As 3$d$ core levels were performed with a pass energy of 20~eV, corresponding to an overall instrument energy resolution of 390~meV full-width-half-maximum. A Shirley-type background has been used to fit and quantify the data. Figure 4a shows the XPS results for an as-grown untreated test sample against a treated test sample with an (3.2 $\pm$ 0.2) nm coating of alumina (Al$_2$O$_3$). Peaks near 42~eV are due to As 3$d$ electrons, while those near 19~eV are due to Ga 3$d$ electrons. In the untreated sample, both As and Ga peaks contain strong satellite features due to surface oxides (As$_2$O$_3$, As$_2$O$_5$, Ga$_2$O$_3$), which contribute $>$ 20\% of the spectral weight and correspond to a native oxide thickness of approx.\ 1.2~nm, in good agreement with the ellipsometry results discussed below. On the other hand, the ALD treated sample shows complete suppression of the As- and Ga-oxide peaks, in agreement with the passivation of the surface by the Al$_2$O$_3$ layer. It has previously been noted that ALD growth of Al$_2$O$_3$ on GaAs replaces the native oxide during the early stages of growth \cite{huang2005surface}. However, both samples also contain peaks due to elemental As and Ga, which appear at slightly higher binding energies (see Figure 4a).

\begin{figure}[ht]
\centering
\begin{subfigure}
         \centering
         \includegraphics[width=0.98\linewidth]{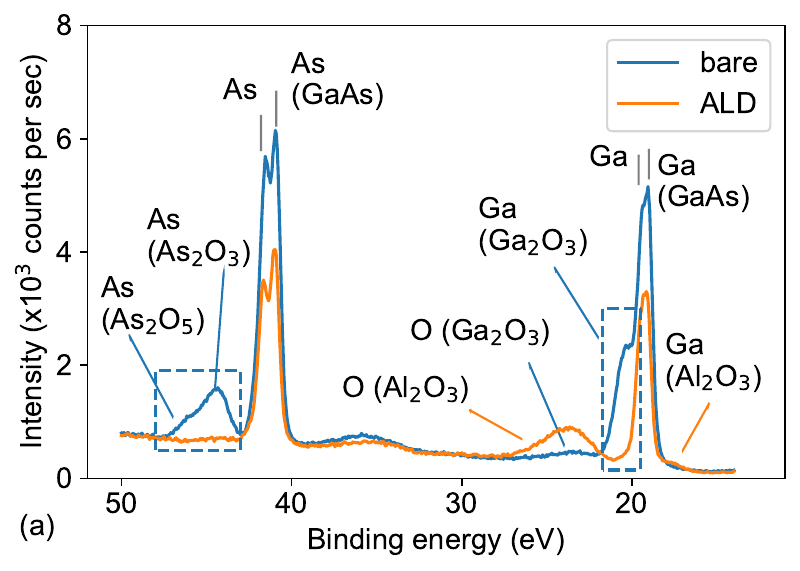}
         \end{subfigure}
         \begin{subfigure}
         \centering
         \includegraphics[width=0.98\linewidth]{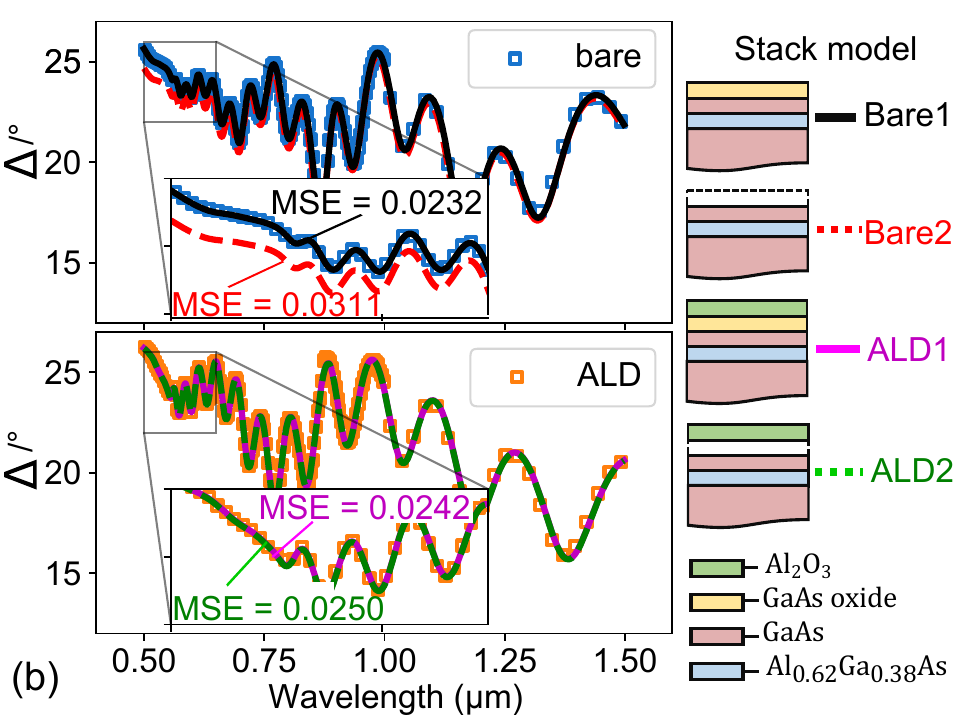}
         \end{subfigure}
\caption{(a) XPS surface chemistry analysis for ALD treated (orange lines) and bare untreated (blue lines) (b) Ellipsometry data of measured polarization phase difference (${\Delta}$) for a bare GaAs sample (top left) and ALD-treated GaAs sample (bottom left). (insets) ${\Delta}$ plots in more detail over the range $\lambda$ = 0.5 \textmu m to 0.65 \textmu m and MSE values. (right) Stack models with and without native oxide layer used to fit the measured ${\Delta}$.}
\label{fig4}
\end{figure}

We note that the data cannot be accurately fitted without including both of these components. Elemental As is well known to be present at the surface of GaAs in the form of As-As dimers, even during Al$_2$O$_3$ passivation \cite{hinkle2008gaas}, and our high-resolution measurements show this is also the case for elemental Ga too. Both As-As and Ga-Ga bonds are predicted to contribute states within the gap \cite{lin2011defect}, which can contribute to residual surface absorption. Overall, the XPS results indicate that ALD growth of Al$_2$O$_3$ on the surface of GaAs successfully passivates the surface from oxide formation and reduces surface absorption. However, As-As and Ga-Ga dimers remain at the interface and removing them using hydrochloric acid \cite{stanton2020efficient} before ALD will improve performance further.

The results of XPS are verified using spectroscopic ellipsometry (SE) which provides an in-fab tool for sensitive surface monitoring \cite{rauch2022model}. A bare untreated sample is measured by SE analyzing the original layer structure, with optical constants for native GaAs oxide, GaAs and AlGaAs (see supplementary material Figure \ref{subfig4}). The sample is then etched, cleaned and coated with alumina using ALD. The mean squared error (MSE) between the measured data and the layer fit serves as a metric to evaluate fit fidelity. As shown in Figure 4b (top), the native GaAs oxide layer can be accurately captured using SE, which estimates the native oxide thickness to be $\approx$ 1.35 nm, in reasonable agreement with the XPS estimate of 1.2 nm (see Appendix B for details). Without the oxide layer, the MSE increases considerably and the fit deviates significantly at shorter wavelengths, as shown in the figure inset. Figure 4b (bottom) shows SE fitting of the ALD-coated sample. From ALD model 1 (ALD1), the fitted native oxide thickness is close to 0 nm and a fitting model (ALD2) without the native oxide layer has no effect on the MSE, although fitting layers with thickness close to 0 increases the error bars significantly, see Appendix B. If we rely on the mean value for layer thickness, the SE fits do indicate that the native oxide is completely removed during the etching/cleaning and ALD deposition process, although this claim can only be made with the supporting XPS evidence in mind. 

\section{Prospects for further improvement}

Moving forward, the two main sources of loss that need to be addressed are the residual bulk absorption in GaAs that can be improved through reducing the background oxygen content in the MBE growth chamber, and removing the As-As dimers on the surface through an HCl clean before surface passivation \cite{stanton2020efficient}. Addressing these will pave the way for GaAs PICs to achieve optical loss performance comparable to their silicon photonics counterparts.

\section{\label{sec:level1}Acknowledgements}

The authors acknowledge the use of the University of Bristol Ultraquiet NanoESCA Laboratory (BrUNEL). This work received funding support from the UK's Engineering and Physical Sciences Research Council (EP/V052179/1) and the European Research Council (SBS3-5, 758843). Nanofabrication was carried out using equipment funded by an EPSRC captial equipment grant (EP/N015126/1). The GaAs wafers were sourced from the UK's national epitaxy facility (EP/X015300/1). We would like to thank Pisu Jiang, Imad Faruque, Laurent Kling, Ankur Khurana, and Jorge Monroy Ruz for helpful discussions and suggestions.

\appendix{}
\section{Characterizing optical loss in microring resonators} 
For the low intracavity optical powers over which the histogram data in Figure 3a were measured ($P_{cav}$ = 1.3 $\pm$ 0.4 mW), the Lorentzian transmission model provides a good fit to the measured data i.e. $R^2 > $ 0.99. However, as can be seen in Figure 2a, at higher intracavity powers ($P_{cav}\approx$ 3 mW) the model fit begins to deviate from the measured data around the resonance minimum due to the sharp change in the gradient of the transmission. This deviation produces an increasingly large error in the model derived values of the thermo-optic frequency shift, $\delta\omega$, and the intrinsic quality factor, $Q_i$, with increasing intracavity power. To accurately plot the normalised frequency shift over a large range of powers, including $P_{cav} >$ 3 mW, as in Figure 3b, the measured data is analysed using additional steps noted below. 

To calculate the frequency shift, $\delta\omega = \omega_0 - \omega_{min}$, the centre frequency, $\omega_0$, is still taken from the Lorentzian model fit, but the frequency of the transmission minimum, $\omega_{min}$, is taken from the lowest value in the measured data, not the model because of the fit discrepancy referred to above. In addition to this, as the expression for $P_{cav}$ depends on the finesse ($F$) of the resonance, which is in turn derived from $Q_i$, it is necessary to use an alternative method for calculating $Q_i$ to avoid unwanted parameter dependencies. Here we use the relationship \cite{bogaerts2012silicon} between the extinction ratio ($ER$) of the resonance, $ER = T_{max}/T_{min}$, for normalized transmission ($T_{max}$ = 1), and extract the round trip amplitude loss coefficient:
\begin{equation} \label{eq:A1}
a =  
	\frac{-\sqrt{ER}{(r^2-1)}+ER\cdot{r}-r}{ER-r^2} 
  \end{equation}
where $a$ accounts for the reduction in the field amplitude during a single round trip of the MRR cavity due to propagation loss and $r$ is the through coupling coefficient that is used to model the waveguide cavity interaction, see Figure 5. The value of $r$ is obtained using the Lorentzian model at powers below 2.5 mW, using $r^2 = \exp{-(n_g/c\cdot{\kappa_c}\cdot{L})}$, where $n_g$ is the group refractive index of the MRR mode, $L$ is the round trip length of the ring and $\kappa_c$ is the coupling rate fitting parameter of the Lorentzian model. Here we assume that non-linear absorption is negligible over the range of powers measured and $r$ can be treated as a constant over this range. From the value of $a$ calculated using equation A1, $Q_i$ can be found from the relationship between $a$ and the intrinsic loss per unit length, $\alpha_i = -2\ln{(a)}/L$.   

\begin{figure}[ht]
\centering
\includegraphics[width=1\linewidth]{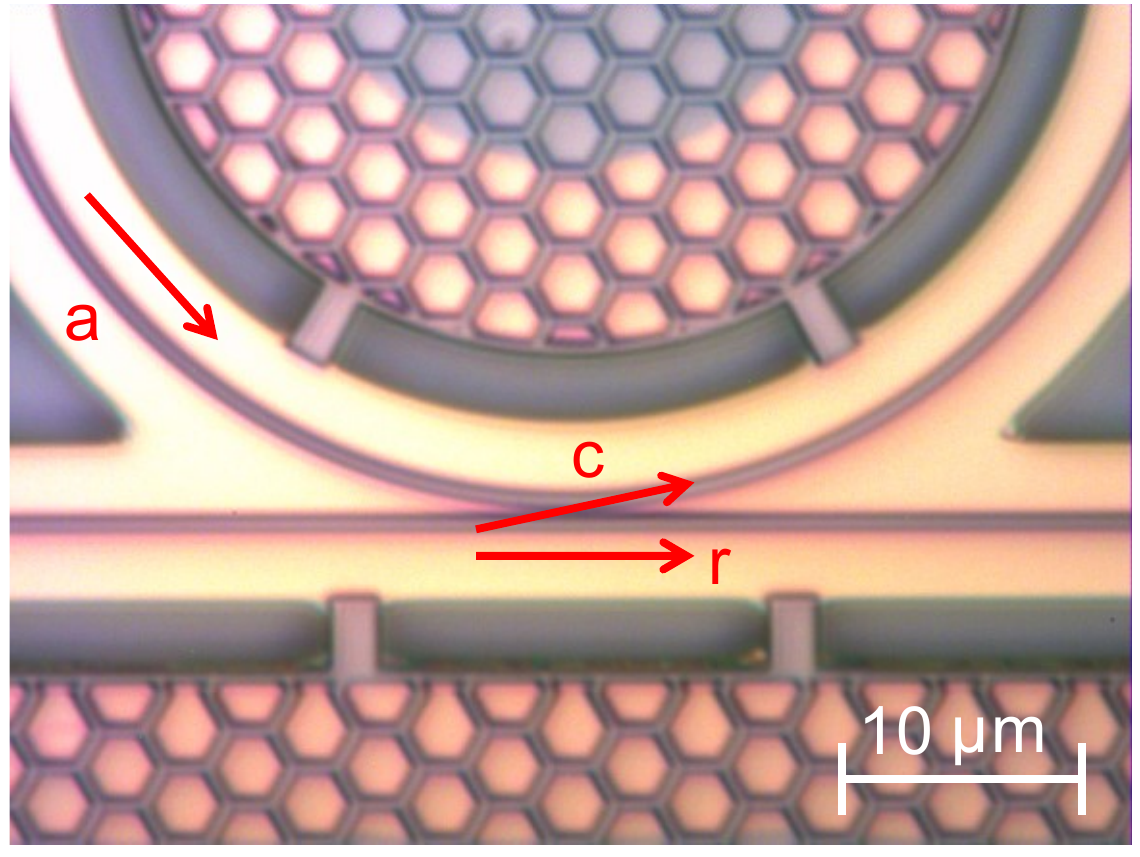}
\caption{Microscope image of the point coupler region of a suspended GaAs micro-ring resonator indicating the E-field amplitude coefficients for $a$ the round trip transmission loss and $c$ and $r$ the MRR/bus waveguide cross-coupling and through coupling coefficients respectively.}
\label{subfig1}
\end{figure}
\begin{figure}[ht]
\centering
     \includegraphics[width=1\linewidth]{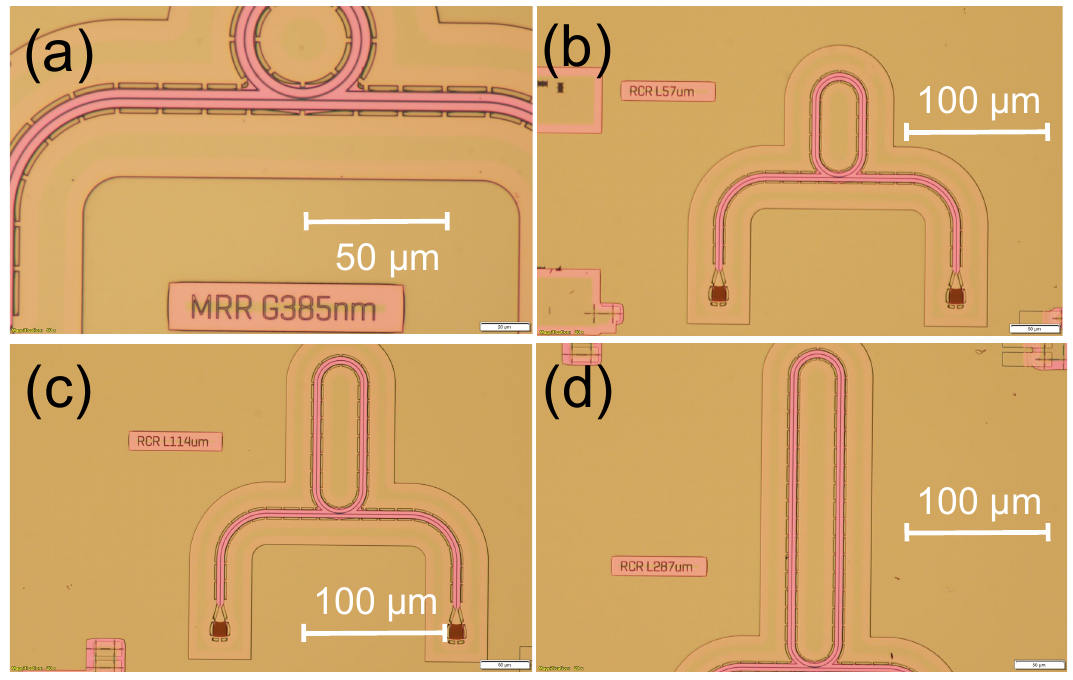}
\caption{(a)-(d) Microscope images of ring and racetrack resonator geometries with identical coupling region and varying cavity lengths to determine the operating regime of our devices.}
\label{subfig2}
\end{figure}
\begin{figure}[!h]
\centering
     \includegraphics[width=1\linewidth]{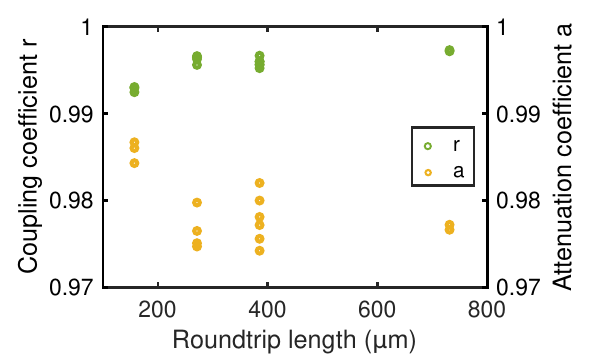}
     \caption{Extracted coupling coefficient ${r}$ and attenuation coefficient ${a}$  as a function of roundtrip length of the ring resonator. The coupling geometry is identical in all devices. }
 \label{subfig3}
\end{figure}
That the microring resonator is operating in the under-coupled regime is determined by measuring the ring quality factor and extinction ratio as a function of roundtrip length ${L}$ with a fixed coupling geometry, as shown in Figure \ref{subfig2}. The ${r}$ and ${a}$ coefficients can be extracted with measured $Q$ and $ER$. Mathematically this solves out two sets of results, one of which shows a consistent value of all roundtrip lengths. We determine the less varying set of results to be ${r}$ because the coupling region is designed to be identical for all cavity lengths. The other set is transmission coefficient ${a}$ which decreases along with the increasing roundtrip length. As shown in Figure \ref{subfig3}, $r>a$  for all the devices considered in this work, and hence the resonators are working in the under-coupled regime even after ALD deposition.
% ============== Table ==========================
\begin{table*}
\begin{adjustbox}{width=\linewidth,center}
\renewcommand{\arraystretch}{1.2}   %set row height
\begin{tabular}{cccccccc}
 \toprule
Sample & VASE stack model & Al${_2}$O${_3}$ & GaAs oxide & GaAs & Al${_x}$Ga${_{1-x}}$As & ${x}$ & MSE\\
\midrule
\multirow{2}{*}{Bare GaAs} & \multicolumn{1}{c}{Bare 1}   & \multicolumn{1}{c}{-} & \multicolumn{1}{c}{1.35 ${\pm}$ 0.28} & \multicolumn{1}{c}{368.61 ${\pm}$ 0.09} & \multicolumn{1}{c}{1048.04 ${\pm}$ 0.21} & \multicolumn{1}{c}{0.624 ${\pm}$ 0.001} & \multicolumn{1}{c}{0.0283}\\\cline{2-8}
  & \multicolumn{1}{c}{Bare 2}& \multicolumn{1}{c}{-} & \multicolumn{1}{c}{-} & \multicolumn{1}{c}{369.31 ${\pm}$ 0.11} & \multicolumn{1}{c}{1044.73 ${\pm}$ 0.28} & \multicolumn{1}{c}{0.623 ${\pm}$ 0.001} & \multicolumn{1}{c}{0.0311}\\
  \hline
\multirow{2}{*}{ALD treated} & \multicolumn{1}{c}{ALD 1} & \multicolumn{1}{c}{7.04 ${\pm}$ 12.84} & \multicolumn{1}{c}{0.27 ${\pm}$ 11.94} & \multicolumn{1}{c}{210.87 ${\pm}$ 0.17}& \multicolumn{1}{c}{1048.27 ${\pm}$ 0.18} & \multicolumn{1}{c}{0.629 ${\pm}$ 0.001}& \multicolumn{1}{c}{0.0242}\\\cline{2-8}
& \multicolumn{1}{c}{ALD 2} & \multicolumn{1}{c}{7.47 ${\pm}$ 0.15} & \multicolumn{1}{c}{-} & \multicolumn{1}{c}{211.19 ${\pm}$ 0.06} & \multicolumn{1}{c}{1048.75 ${\pm}$ 0.18} & \multicolumn{1}{c}{0.629 ${\pm}$ 0.001} & \multicolumn{1}{c}{0.0264}\\
\bottomrule
\end{tabular}
\end{adjustbox}
\label{Tbl1}
\caption{ Summary of fitted layer thickness (in nm) of the bare GaAs sample and the ALD treated sample, after etching, using different stack models specified in Figure 4b. The measurement and model fitting are both accomplished using the CompleteEASE$^{\text{TM}}$ software. The overall mean square error (MSE) is used as a metric to determine fit fidelity.}
\end{table*}
% ============== Table ==========================

\section{Spectroscopic Ellipsometry layer fits}
\maketitle

 Spectroscopic ellipsometry (SE) is a standard tool for determining thin film thicknesses. Recently, it was shown \cite{rauch2022model} that the technique can accurately predict native oxide thicknesses provided the fit parameters are carefully controlled and an independent method is used to verify the results. In this work, we primarily use SE as an in-fab tool verify layer thicknesses in-process. The results reported here should be primarily interpreted as a confirmation of the more accurate XPS results.
 
 The SE spectra of the phase difference ($\Delta$) and the amplitude ratio ($\psi$) are measured over a wavelength range of 500 nm-1500 nm at room temperature, using a Woollam v-VASE spectroscopic ellipsometer. Only the $\Delta$ values are depicted in the main text. The refractive index ($n,k$) models of the optical materials in the GaAs device stack that have been used for fitting the ellipsometry data are plotted in Figure 8. The values for the estimated layer thicknesses for the different stacks in Figure 4b, along with the fit errors for each layer, and the overall mean squared error (MSE) are shown in Table I. 
 
 Our main goal with using SE was to confirm the presence of native oxide on a bare substrate and its absence after ALD treatment. For the bare GaAs sample (rows 1 and 2) in the table, adding a GaAs surface oxide layer both reduces the overall MSE and also provides a native oxide thickness of 1.35 nm that is in close agreement with the XPS determined value of 1.2 nm, as discussed in the main text. For the ALD treated sample, the overall MSE is unchanged with and without the presence of the native oxide layer. Moreover, forcing a fit to an oxide layer (row 3) shows its thickness close to zero, which also indicates that the native oxide is removed during the dry etching and ALD treatment process. We would like to note here that the large error bars in row 3 for the alumina and oxide layers are primarily due to trying to forcefully fit a layer with near-zero thickness. One way to see this is to compare the error bars in row 4 where the oxide layer is removed, but the overall MSE is unchanged. Here, the layer thickness errors are minimal.  
\begin{figure}[H]
\centering
    \includegraphics[width = 1\linewidth]{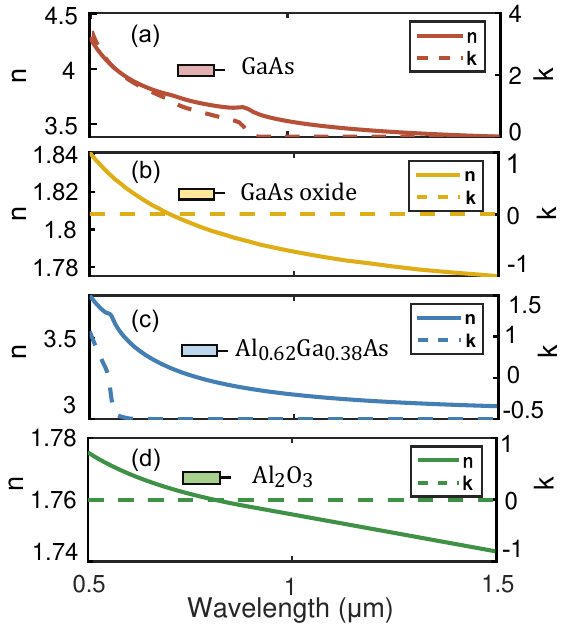}
\caption{Refractive index (n,k) data for the different layers used to fit the ellipsometry data. (a) GaAs \cite{snyder1992study} (b) GaAs oxide \cite{zollner1993model} (c) AlGaAs \cite{herzinger1996studies} (d) Al$_2$O$_3$ \cite{KUNKUO2002675}}
\label{subfig4}
\end{figure}
\bibliography{References}% Produces the bibliography via BibTeX.

\end{document}